\documentclass[11pt]{revtex4}
\usepackage{amssymb,epsf}
\usepackage{latexsym}
\begin{document}
\title{Extremal Myers-Perry black holes coupled to
Born-Infeld electrodynamics in odd dimensions}
\author{Masoud Allaverdizadeh$^{1}$\footnote{masoud.alahverdi@gmail.com},
Seyed H. Hendi $^{2,3}$\footnote{hendi@shirazu.ac.ir}, Jos\'e P.
S. Lemos$^{1}$\footnote{joselemos@ist.utl.pt}, Ahmad
Sheykhi$^{2,3}$\footnote{asheykhi@shirazu.ac.ir}}
\address{ $^1$  Centro Multidisciplinar de Astrof\'{\i}sica -
CENTRA, Departamento de F\'{\i}sica,
Instituto Superior T\'ecnico - IST, Universidade T\'ecnica de Lisboa -
Av. Rovisco Pais 1, 1049-001 Lisboa, Portugal. \\
          $^2$  Physics Department and Biruni Observatory,
College of Sciences, Shiraz University, Shiraz 71454, Iran\\
          $^3$ Center for Excellence in Astronomy and 
Astrophysics (CEAA-RIAAM), Maragha P. O. Box 55134-441, Iran}

\begin{abstract}
\vspace*{1.5cm}
\centerline{\bf Abstract}
\vspace*{1cm}

Employing higher order perturbation theory, we find a new class of
perturbative extremal rotating black hole solutions with Born-Infeld
electric charge in odd $D$-dimensional spacetime.  The seed solution
is an odd dimensional extremal Myers-Perry black hole with equal
angular momenta to which a perturbative nonlinear electric Born-Infeld
field charge $q$ is added maintaining the extremality condition.  The
perturbations are performed up to third-order. We also study some
physical properties of these black holes. In particular, it is shown
that the values of the gyromagnetic ratio of the black holes are
midified by the perturbative parameter $q$ and the Born-Infeld
parameter $\beta$.

\end{abstract}
 \maketitle
\section{Introduction\label{Intro}}

Finding rotating black holes solutions in higher dimensions is a
difficult task due to the size and complexity of the equations. For
this reason the number of analytic solutions in closed form is very
limited. An exception is pure general relativity, where there are
exact $D$-dimensional rotating black holes. Indeed, the generalization
of the Kerr metric to higher $D$ dimensions was performed by Myers and
Perry \cite{Myer}. These Myers-Perry black holes in general possess
$N=[(D-1)/2]$ independent angular momenta as stated in \cite{Myer},
where $[X]$ denotes the integer part of $X$. This, in turn, implies,
that $D$-dimensional rotating black hole solutions fall into two
classes, namely, even-$D$ and odd-$D$. The insertion of a cosmological
constant also yields an exact rotating solution in $D$-dimensions
\cite{cosmo}.

The inclusion of any other charge into the solutions, be it electric,
magnetic, or dilatonic, to name a few, obliges the introduction of
alternative techniques to study rotating black holes in higher
dimensions.  There are then two opposed situations that can be dealt
with some ease: slow rotation and rotation in the extremal regime.
For slow rotation, charged black holes with a single rotation
parameter in higher dimensions have been studied perturbatively in
\cite{Aliev2,Aliev3,Hendi,Aliev4,kunz1}, and numerically in
\cite{Kunz2} for asymptotically flat solutions and in \cite{Kunz3} for
asymptotically anti-de Sitter solutions.  The incorporation
of a dilaton
coupling in this slow rotation regime in four dimensions has been
done in \cite{HorneShiraishi} and it is also
possible to find the corresponding higher-dimensional solutions.

For the rotating solutions in the extremal regime there are
perturbative methods that work in odd $D$-dimensions when one includes
some type of charge.  The correctness of these perturbative 
solutions can be performed 
by making use of a Smarr-type formula for
black holes in $D$ dimensions \cite{Gauntlett}.
Employing perturbation theory with the electric
charge as the perturbation parameter, charged rotating
Einstein-Maxwell black holes have been constructed in five dimensions
\cite{Navarro}. This perturbative method was also applied to obtain
extremal Einstein-Maxwell black holes with equal magnitude angular
momenta in odd dimensions \cite{Allahverdi1}. These solutions have
then been generalized by including a scalar dilaton field
\cite{Allahverdi2,Allahverdi3}.  One can also deal with Born-Infeld
electric charge, indeed, extremal rotating Einstein-Born-Infeld black
holes in five-dimensional spacetime have been studied
\cite{Allahverdi4}.

In this paper, we want to generalize the five-dimensional studied
performed in \cite{Allahverdi4} to any odd $D$ dimension, $5\leq
D<\infty$. The odd-dimensional case, in opposition to the
even-dimensional case, can be treated explicitly perturbatively when
the $N$ angular momenta of the black hole have all equal magnitude, as
the resulting system of field equations simplifies remarkably.  Using
a prescribed perturbative method we are able to find extremal rotating
Einstein-Born-Infeld black holes.  We start from the extremal
Myers-Perry black holes with equal $N=[(D-1)/2]$ angular momenta
\cite{Myer}, then we evaluate the perturbative series up to third
order in the electric charge parameter $q$, and finally we study the
physical properties of these black holes. In particular, we analyze
the effects of the perturbative parameter $q$ and the Born-Infeld
parameter $\beta$ on the gyromagnetic ratio of these fast rotating
black holes.  The structure of this paper is the following: In section
\ref{first}, the field equations of the nonlinear Born-Infeld theory
in Einstein gravity are displayed and a new class of perturbative
charged rotating solutions in odd dimensions is obtained. In section
\ref{pq}, the physical quantities of the solutions are calculated and
their properties discussed.  In Sec.~\ref{MassForm} the mass formula
for these black holes is presented.  In section \ref{sumcon} we draw
some conclusions. The formulas for the metric and the gauge potential
in $D$ dimensions are given in the Appendix.

\section{METRIC AND BASIC EQUATIONS \label{first}}

Our departure point is the Einstein-Hilbert action coupled to the
Born-Infeld nonlinear gauge field in $D$ dimensions
\begin{eqnarray}
S &=&\int dx^{D}\sqrt{-g}\left( \frac{R}{16 \pi G_{D}}\text{
}+L(F)\right),  \label{Lag}
\end{eqnarray}
where $G_{D}$ is the Newton constant in $D$ dimensions,
$g$ is the determinant of the $D$-dimensional metric
$g_{\mu \nu }$,
${R}$ is the
Ricci curvature scalar. $L(F)$ is the Lagrangian of the nonlinear
Born-Infeld gauge field given by
\begin{eqnarray}
L(F) &=&4\beta^{2}\left(1-\sqrt{1+\frac{F}{2 \beta^{2}}}\right),
\label{Born-Infeld}
\end{eqnarray}
where, $F=F^{\mu \nu }F_{\mu \nu
}$, $F_{\mu \nu }=\partial _{\mu }A_{\nu }-\partial
_{\nu }A_{\mu }$ is the electromagnetic field tensor, $A_{\mu }$
is the electromagnetic vector potential, and $\beta$ is the
Born-Infeld parameter. In the limit $\beta
\rightarrow \infty $, $L(F)$ reduces to the Lagrangian of the
standard Maxwell field, $L(F)=F$. By varying the action with
respect to the gravitational field $g_{\mu \nu }$ and the gauge
field $A_{\mu }$ one obtains the
equations of motion for these fields.
For the gravitational field the
equations are
\begin{equation}
G_{\mu \nu }=
\frac{1}{2}g_{\mu \nu } L(F)+\frac{2 F_{\mu \eta }F_{\nu }^{\text{
}\eta }}{\sqrt{1+\frac{F}{2 \beta^{2}}}}\,,
\label{FE1}
\end{equation}
where $G_{\mu \nu }=R_{\mu \nu }-\frac{1}{2}g_{\mu \nu } R$
is the Einstein tensor formed out from the Ricci tensor $R_{\mu \nu }$ 
and scalar $R$. For the Born-Infeld electromagnetic field one finds
the following equations
\begin{equation}
\partial_{\mu}{\left(\frac{\sqrt{-g}F^{\mu \nu }}
{\sqrt{1+\frac{F}{2 \beta^{2}}}}
\right)}=0\,. \label{FE2}
\end{equation}
Here, we consider extremal charged rotating
black hole solutions of the above field equations in odd
dimensions through a perturbative
method. To obtain such perturbative charged generalizations of the
$D$-dimensional Myers-Perry solutions \cite{Myer}, we employ the
following parametrization for the metric
\cite{Allahverdi1,Allahverdi2,Allahverdi3},
\begin{eqnarray}\label{metric1}
\phantom{a=a}\nonumber \\
ds^2 &=&g_{tt}dt^2+\frac{dr^2}{W} + r^2
\left[\sum^{N-1}_{i=1}\left(\prod^{i-1}_{j=0}
\cos^{2}\theta_{j}\right)d\theta^{2}_{i}+
\sum^{N}_{k=1}\left(\prod^{k-1}_{l=0}\cos^{2}
\theta_{l}\right)\sin^{2}\theta_{k}d\varphi^{2}_{k}\right]
\nonumber \\[10pt]
&+&V\left[\sum^{N}_{k=1}\left(\prod^{k-1}_{l=0}
\cos^{2}\theta_{l}\right)\sin^{2}\theta_{k}
\varepsilon_{k}d\varphi_{k}\right]^{2}-2B\sum^{N}_{k=1}
\left(\prod^{k-1}_{l=0}\cos^{2}\theta_{l}\right)
\sin^{2}\theta_{k}\varepsilon_{k}d\varphi_{k}dt \,  ,
\phantom{a=a}\nonumber \\
\end{eqnarray}
where $\theta_{0}\equiv0$, $\theta_{i}\in[0,\pi/2]$
for $i=1,...,N-1$, $\theta_{N}\equiv \pi/2$,
$\varphi_{k}\in[0,2\pi]$ for $k=1,...,N$,
and $\varepsilon_{k}=\pm1$ denotes the sense of rotation
in the $k$-th orthogonal plane of rotation, and the metric
functions 
$g_{tt}$, $W$, $V$, and $B$
depend only on the radial coordinate $r$.
An adequate parametrization for the gauge potential is given by
\begin{eqnarray}\label{A1}
A_{\mu}dx^{\mu} &=& a_{t} dt
+a_{\varphi}\sum^{N}_{k=1}\left(\prod^{k-1}_{l=0}\cos^{2}\theta_{l}\right)
\sin^{2}\theta_{k}\varepsilon_{k}d\varphi_{k} \, .
\end{eqnarray}
where the gauge potential functions $a_{t}$ and $a_{\varphi}$,
depend only on the radial coordinate $r$.

We now consider perturbations around the Myers-Perry solution, with a
Born-Infeld electric charge $q$ as the perturbative parameter.  Taking
into account the seed solution and the symmetry with respect to charge
reversal, the functions for the metric and gauge potential take the
form
\begin{eqnarray}\label{gtt}
g_{tt} = -1+\frac{2\hat{M}}{r^{D-3}}+q^{2}g^{(2)}_{tt}
+O(q^{4}) \ ,
\end{eqnarray}
\begin{eqnarray}\label{W}
W =1-\frac{2\hat{M}}{r^{D-3}}+\frac{2\hat{J}^{2}}{\hat{M}r^{D-1}}
+q^{2}W^{(2)}+O(q^{4}) \ ,
\end{eqnarray}
\begin{eqnarray}\label{N}
V = \frac{2\hat{J}^{2}}{\hat{M}r^{D-3}}+q^{2}V^{(2)}
+O(q^{4}) \ ,
\end{eqnarray}
\begin{eqnarray}\label{B}
B = \frac{2\hat{J}}{r^{D-3}}+q^{2}B^{(2)}
+O(q^{4}) \ ,
\end{eqnarray}
\begin{eqnarray}\label{a0}
a_{t} = q a^{(1)}_{t}
+q^{3} a^{(3)}_{t}+O(q^{5}) \ ,
\end{eqnarray}
\begin{eqnarray}\label{avarphi}
a_{\varphi} = q a^{(1)}_{\varphi}
+q^{3} a^{(3)}_{\varphi}+O(q^{5}) \ ,
\end{eqnarray}
Here $g^{(2)}_{tt}$ is
a second-order perturbative term with the
other perturbative terms defined similarly.
The quantities
$\hat{M}$ and
$\hat{J}$ are integration
constants related to the mass and angular momentum, respectively.

It is important to be able to fix the integration
constants. First one fixes the angular momenta at any perturbative
order and then imposes the extremality condition in all orders.
We also assume
asymptotic flatness and regularity at the horizon.
With these assumptions we are able to fix the
constants of integration.

We now introduce a parameter $\nu$ for
the extremal Myers-Perry solutions in $D$ dimensions by
\begin{equation}
\hat{M}=\frac{(D-1)^{\frac{(D-1)}{2}}
}{4(D-3)^{\frac{(D-3)}{2}}}\;\nu^{D-3}\ ,
\quad\quad
\hat{J}=\frac{(D-1)^{\frac{(D-1)}{2}}
}{4(D-3)^{\frac{(D-3)}{2}}}\;\nu^{D-2}\ .
\label{nu_gen}
\end{equation}
Inserting the metric, Eq.~(\ref{metric1}), and the gauge potential,
Eq.~(\ref{A1}), together with the perturbation expansions,
Eqs.~(\ref{gtt})-(\ref{avarphi}), into the field equations
Eqs.~(\ref{FE1})-(\ref{FE2}), we can solve these equations order by
order. The perturbative expansions for the metric and the gauge
potential functions are exhibited in Appendix A for generic values of
the Born-Infeld parameter $\beta$. One may note that in the Maxwell
limit, $\beta \longrightarrow \infty $, these perturbative solutions
reduce to the odd dimensional perturbative charged rotating black
holes in Einstein-Maxwell theory presented in \cite{Allahverdi1} and
the accuracy of these solutions can be cheked by using Smarr's
formula
\cite{Gauntlett}. 
Also, for the case $D=5$ one finds the equations found in
\cite{Allahverdi4}.

\section{Physical Quantities\label{pq}}

Writing the  asymptotic behavior of the metric and the gauge potential
one can extract the mass $M$, the equal-magnitude angular momenta $|J_i|=J$,
the electric charge $Q$, and the magnetic moments $\mu_{\rm mag}$.
Indeed, one can set the asymptotic behavior as
\cite{Allahverdi1,Allahverdi2,Allahverdi3,Allahverdi4}
\begin{eqnarray}\label{quantities}
&&g_{tt}=-1+\frac{\tilde{M}}{r^{D-3}}+... \ ,  \quad
B=\frac{2\tilde{J}}{r^{D-3}}+... \ ,
\nonumber\\
&&a_{t}=\frac{\tilde{Q}}{r^{D-3}}+... \ ,  \quad
a_{\varphi}=\frac{\tilde{\mu}_{\rm mag}}{r^{D-3}}+... \ ,
\end{eqnarray}
where $\tilde{M}$,
$\tilde{J}$,
$\tilde{Q}$, and $\tilde{\mu}_{\rm mag}$ are parameters 
with dimensions of mass, angular momentum, charge, 
and magnetic momentum, respectively, and we have put
$\tilde{Q}\equiv q$. These parameters are related to 
the mass $M$, angular momentum $J$, electric charge $Q$,
and magnetic momentum ${\mu}_{\rm mag}$ through 
\begin{eqnarray}\label{quantities1}
\tilde{M}&=&\frac{16\pi G_{D}}{(D-2)A(S^{D-2})}\,M \ ,
\nonumber \\
\tilde{J}&=&\frac{4\pi G_{D}}{A(S^{D-2})}\,J \ ,\nonumber \\
\tilde{Q}&=&\frac{4\pi G_{D}}{(D-3)A(S^{D-2})}\,Q \ ,
\nonumber \\
\tilde{\mu}_{\rm mag}&=&\frac{4\pi G_{D}}{(D-3)A(S^{D-2})}\,\mu_{\rm mag} \ ,
\end{eqnarray}
and where
$A(S^{D-2})$ is the area of the unit $(D-2)$-sphere. Comparing the
above expansions to the asymptotic behavior of the solution,
Eqs.~(\ref{gg_{tt}})-(\ref{aphiphi}), we obtain
\begin{eqnarray}\label{mass}
M =\frac{A(S^{D-2})}{64\pi G_{D}}\left[
\frac{2\nu^{2(D-3)}(D-2)(D-1)^{D-1}+16q^{2}(D-3)^{D-2}}{
\nu^{D-3}(D-1)^{\frac{D-1}{2}}(D-3)^{\frac{D-3}{2}}}
\right]+O(q^{4}) ,
\end{eqnarray}
\begin{eqnarray}\label{charge}
Q =\frac{A(S^{D-2})(D-3)}{4\pi G_{D}}q ,
\end{eqnarray}
\begin{eqnarray}\label{ang-mom}
J =\frac{A(S^{D-2})\nu^{D-2}(D-1)^{
\frac{D-1}{2}}}{16 \pi G_{D}(D-3)^{\frac{D-3}{2}}} ,
\end{eqnarray}
\begin{eqnarray}\label{mag-mom}
\mu_{\rm mag} &=& \frac{A(S^{D-2})(D-3)}{4\pi G_{D}}
\Bigg{\{}q\nu-\frac{4q^3(D-3)^{D-3}(D^3-5D^2+7D-3)}{
\nu^{2D-7}(D-2)^{2}(D-1)^{D}}\nonumber \\
&&-\frac{4q^3(D-3)^{D+1}}{3(3D-5)(3D-7)(D-1)^{D-1}
\beta^{2}\nu^{2D-5}}\Bigg{\}}+O(q^{5})\,.
\end{eqnarray}
The gyromagnetic ratio $g$ is given by
\begin{eqnarray}\label{g}
g &=& \frac{2M\mu_{\rm mag}}{QJ} = (D-2)-\frac{4q^{2}
\left[(D^{2}-4D+3)(D-3)^{D-3}-2(D-2)(D-3)^{D-2}
\right]}{(D-2)(D-1)^{D-1}\nu^{2(D-3)}}\nonumber \\
&&-\frac{4q^2(D-3)^{D+1}(D-2)}{3(3D-5)(3D-7)(D-1)^{D-1}
\beta^{2}\nu^{2D-4}}+O(q^{4})\,.
\end{eqnarray}

\begin{figure}
\epsfxsize=5cm \centerline{\epsffile{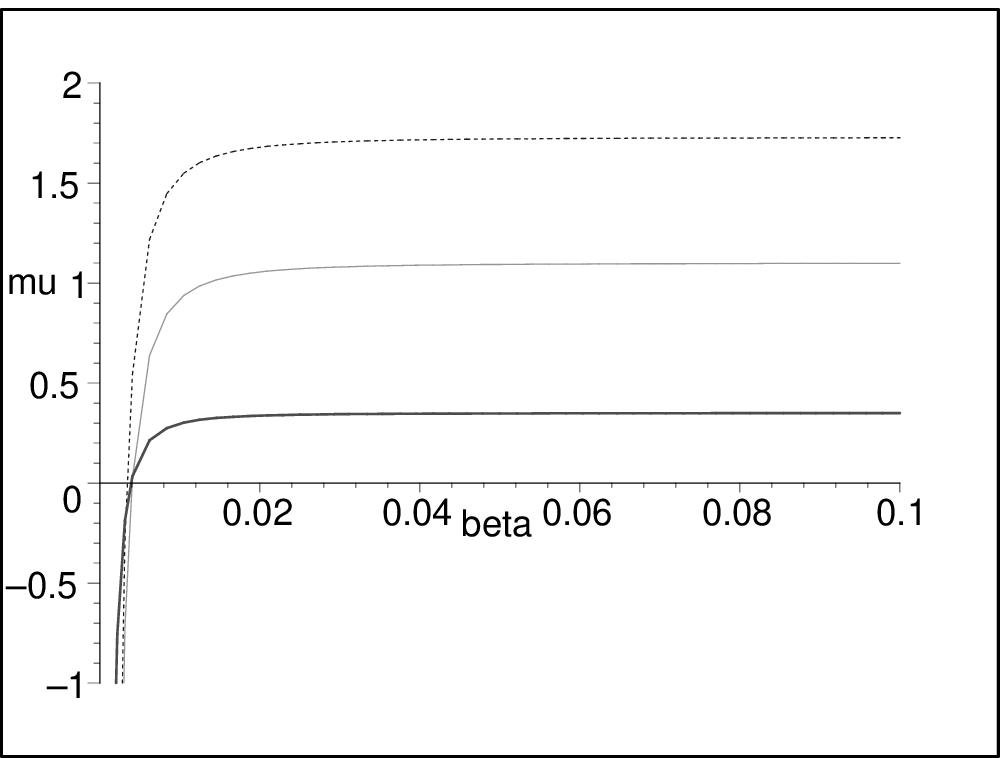}}
\caption{The third-order perturbative values for the magnetic moment
$\mu_{\rm mag}$ versus Born-Infeld parameter $\beta$  for
the perturbative extremal charged rotating black
holes in the Einstein-Born-Infeld theory in 5, 7 and 9 dimensions with
$\nu=1.16$ and $q=0.09$.}
\label{fig2}
\vskip 0.4cm
\epsfxsize=5cm \centerline{\epsffile{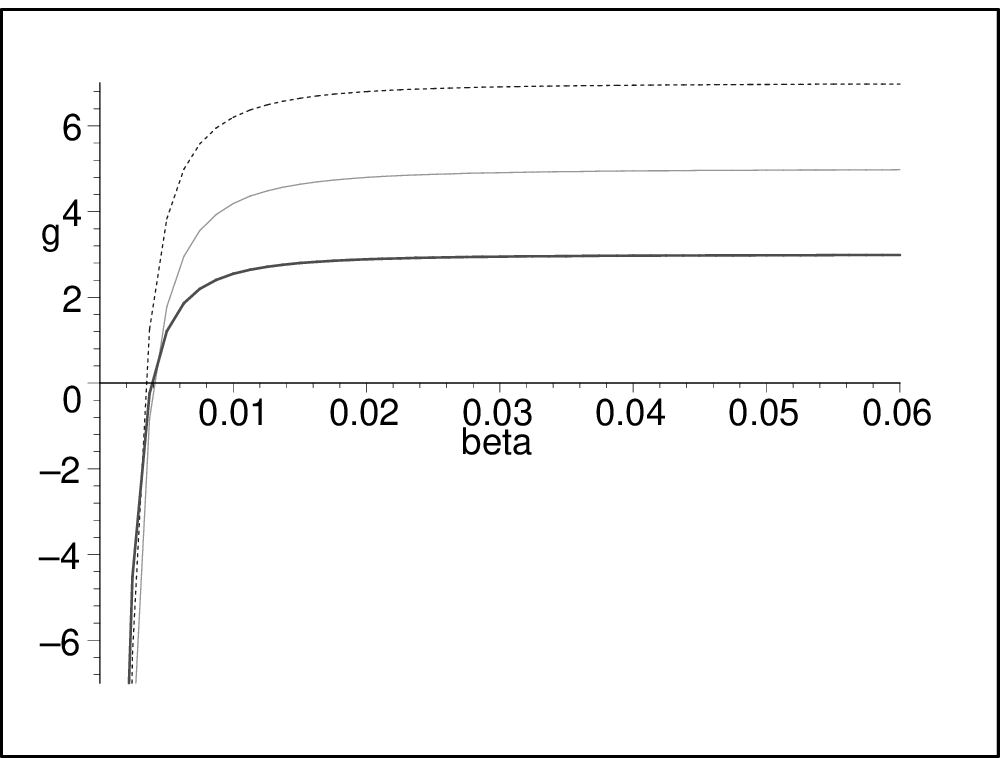}}
\caption{The third-order perturbative values for the gyromagnetic ratio
$g$ versus Born-Infeld parameter $\beta$ for
the perturbative extremal charged rotating black
holes in the Einstein-Born-Infeld theory in 5, 7 and 9 dimensions with
$\nu=1.16$ and $q=0.09$.}
\label{fig3}
\end{figure}

We show the behavior of the magnetic moment $\mu_{\rm mag}$ and
the gyromagnetic ratio $g$ of the perturbative  extremal charged rotating black
holes in the Einstein-Born-Infeld theory versus $\beta$ in Fig.~1
and Fig. 2 respectively. From these figures we find out that the
magnetic moment $\mu_{\rm mag}$ and the gyromagnetic ratio $g$
increase with increasing $\beta$ in any odd dimension. One can
also see
that for some small value of $\beta$ the gyromagnetic ratio and
magnetic moment $\mu_{\rm mag}$ are zero and then become negative. We
speculate that this change of sign comes from significantly different
distributions of charge for high an low values of the Born-Infeld parameter
$\beta$. The interested reader can see \cite{Allahverdi3} for a detailed
discussion about a possible 
interpretation of this sign reversal. In the Maxwell's limit, $\beta
\longrightarrow \infty $, the magnetic moment and the gyromagnetic
ratio reduce to
\begin{eqnarray}\label{mag-momEM}
\mu_{\rm mag} &=& \frac{A(S^{D-2})(D-3)}{4
\pi G_{D}}\Bigg{\{}q\nu-\frac{4q^3(D-3)^{D-3}(D^3-5D^2+
7D-3)}{\nu^{2D-7}(D-2)^{2}(D-1)^{D}}\Bigg{\}}+O(q^{5}),
\end{eqnarray}
\begin{eqnarray}\label{gEM}
g &=& (D-2)-\frac{4q^{2}
\left[(D^{2}-4D+3)(D-3)^{D-3}-2(D-2)(D-3)^{D-2}
\right]}{(D-2)(D-1)^{D-1}
\nu^{2(D-3)}}+O(q^{4}),
\end{eqnarray}
which are exactly the expressions obtained for the odd dimensional
perturbative charged rotating black holes in the
Einstein-Maxwell theory
\cite{Allahverdi1}. The event horizon of these rotating Born-Infeld
black holes is located at
\begin{equation}
r_{\rm H}=\sqrt{\frac{D-1}{D-3}}\nu+\frac{4(D-3)^{
D-\frac{5}{2}}}{(D-2)(D-1)^{D-\frac{3}{2}}\nu^{2D-7}}\,q^2+
O(q^{4}) \, . \label{r_H_odddim}
\end{equation}
Its value does not depend on $\beta$ up to this order.

\section{The mass formula \label{MassForm}}

Now,
the constant horizon angular velocities $|\Omega_i|=\Omega$
can be defined by imposing that the Killing
vector field
\begin{eqnarray}\label{chi}
\chi = \xi+\Omega\sum^{N}_{k=1}\epsilon_{k}\eta_{k},
\end{eqnarray}
is null on and orthogonal to the horizon, with $\chi$ defined
as $\chi=\partial_t$ and
$\eta_{k}=\partial_{\varphi_k}$. The horizon electrostatic potential
$\Psi_{\rm H}$ of these black holes is then given by
\begin{eqnarray}\label{PsiH}
\Psi_{\rm  H} = \left. (a_{t}+\Omega a_{\varphi})
\right|_{r=r_{\rm H}} \, ,
\end{eqnarray}
and the surface gravity $\kappa$ is defined by
\begin{eqnarray}\label{sg1}
\kappa^{2} = \left.
-\frac{1}{2}(\nabla_{\mu}
\chi_{\nu})(\nabla^{\mu}\chi^{\nu})\right|_{r=r_{\rm H}} \, .
\end{eqnarray}
Finally, the horizon angular velocities $|\Omega_i|=\Omega$ and
the horizon area $A_{\rm H}$ are then given by
\begin{eqnarray}\label{Omega}
\Omega = \frac{D-3}{\nu(D-1)}-
\frac{8q^2(D-3)^{D-1}}{\nu^{2D-5}(D-2)(D-1)^{D}}+O(q^{4})\,,
\end{eqnarray}
\begin{eqnarray}\label{AH}
A_{\rm H}=\frac{\sqrt{2}A(S^{D-2})(D-1)^
{\frac{(D-1)}{2}}\nu^{D-2}}{2(D-3)^{\frac{(D-2)}{2}}}+O(q^{4}) \, ,
\end{eqnarray}
respectively.
Taking into account the quantities previously defined, it is
straightforward to see that these black holes satisfy the Smarr mass
formula up to third-order \cite{Gauntlett},
namely $\frac{D-3}{D-2}M =
\frac{\kappa\, A_{H}}{8\pi G_{D}}+
N\Omega J+\frac{D-3}{D-2}\Psi_{\rm H}\,Q$.
Noting that the surface
gravity $\kappa$ vanishes for extremal solutions, one has
in our odd-dimensional case
\begin{eqnarray}\label{smarr}
\frac{D-3}{D-2}M =
N\Omega J+\frac{D-3}{D-2}\Psi_{\rm  H}\,Q.
\end{eqnarray}
For $D=5$ the result in \cite{Allahverdi4} is recovered.

\section{Conclusions\label{sumcon}}

We have constructed a new class of perturbative charged rotating black
hole solutions in higher odd dimensions in the presence of a nonlinear
Born-Infeld gauge field. We have restricted to the case of extremal
black holes with equal angular momenta. These solutions are
asymptotically flat and their horizons have spherical topology. Our
strategy for obtaining these solutions was through a perturbative
method up to the third-order for the perturbative 
charge parameter $q$. We
have started from rotating Myers-Perry black hole solutions in odd
dimensions, and then investigated the effects of adding a charge
parameter $q$ and the Born-Infeld parameter $\beta$ to the
solutions. We have calculated the mass, angular momentum, electric
charge, magnetic moment, gyromagnetic ratio, and horizon radius of
these Born-Infeld black holes.  For large $\beta$ the solutions reduce
to the perturbative rotating Einstein-Maxwell solutions in odd
dimensions \cite{Allahverdi1}, as expected.  Recently, it was shown
that in five dimensions the Born-Infeld parameter $\beta$ may modify
the value of the gyromagnetic ratio relative to the corresponding
Einstein-Maxwell rotating black holes\cite{Allahverdi4}. Here, we
obtain a similar result for odd-dimensional black holes.

\appendix

\section{}

We give the perturbative expressions
for the metric and the gauge potential in the
Einstein-Born-Infeld theory for general odd $D$.
The solutions up to third-order read
\begin{eqnarray}\label{gg_{tt}}
g_{tt} &=& -1+\frac{(D-1)^
{\frac{(D-1)}{2}}\nu^{D-3}}{2(D-3)^{\frac{(D-3)}{2}}r^{D-3}}\nonumber \\
&&-q^{2}\left[\frac{2(D-3)}{(D-2)r^{2(D-3)}}-\frac{4(D-3)^
{\frac{D-1}{2}}}{(D-2)(r\nu)^{D-3}(D-1)^{\frac{D-1}{2}}}\right]+
O(q^{4}) \  ,
\end{eqnarray}
\begin{eqnarray}\label{WW}
W &=&1-\frac{(D-1)^{\frac{(D-1)}{2}}
\nu^{D-3}}{2(D-3)^{\frac{(D-3)}{2}}r^{D-3}}+\frac{(D-1)^{
\frac{(D-1)}{2}}\nu^{D-1}}{2(D-3)^{\frac{(D-3)}{2}}r^{D-1}}
-\frac{2q^{2}}{(D-2)}\Bigg{\{}\frac{2(D-3)^{
\frac{(D-1)}{2}}}{(r\nu)^{D-3}(D-1)^{\frac{(D-1)}{2}}}\nonumber \\
&&+\frac{(D-3)^{\frac{(D-3)}{2}}}{
\nu^{D-5}r^{D-1}(D-1)^{\frac{(D-3)}{2}}}+\frac{(D-5)\nu^{2}}{r^{2(D-2)}}-
\frac{D-3}{r^{2(D-3)}}\Bigg{\}}+O(q^{4}) \ ,
\end{eqnarray}
\begin{eqnarray}\label{VV}
V &=& \frac{(D-1)^{\frac{(D-1)}{2}}\nu^{D-1}}{2(D-3)^{
\frac{(D-3)}{2}}r^{D-3}}\nonumber \\
&&-q^{2}\left[\frac{2(D-3)\nu^2}{(D-2)r^{2(D-3)}}+\frac{4(D-3)^{
\frac{D-3}{2}}}{(D-2)(D-1)^{
\frac{D-3}{2}}\nu^{D-5}r^{D-3}}\right]+O(q^{4}) \ ,
\end{eqnarray}
\begin{eqnarray}\label{BB}
B = \frac{(D-1)^{\frac{(D-1)}{2}}
\nu^{D-2}}{2(D-3)^{\frac{(D-3)}{2}}r^{D-3}}-
\frac{2\nu(D-3)q^{2}}{(D-2)r^{2(D-3)}}+O(q^{4}) \ ,
\end{eqnarray}
\begin{eqnarray}\label{a00}
a_{t} &=& \frac{q}{r^{D-3}}+q^3\Bigg{\{} \int \! \Bigg{\{}
\int \!S_{{4}} \left(
 \left( \frac{D-3}{2} \right) {r}^{2}- \left(
\frac{D-1}{2} \right) {\nu}^{2}
 \right) ^{2}S_{{1}}\int \!{\frac {{r}^{D-2}}{{S_{{4}}}^{2}
\left( \left( \frac{D-3}{2} \right) {r}^{2}-
\left( \frac{D-1}{2} \right) {\nu}^{2}
 \right) ^{4}}}{dr}\nonumber \\
&&\left( {r}^{2-3\,D} \right){dr}{r}^{3\,D-3}+
S_{{2}}\int \!{\frac {{r}^{D-2}}{{S_{{4}}}^{2}
\left( \left( \frac{D-3}{2}
 \right) {r}^{2}-\left( \frac{D-1}{2}
\right) {\nu}^{2} \right) ^{4}}}{dr}
+{\frac {{r}^{3\,D-3} \left( D-1 \right) ^{\frac{D+3}{2}} {
\nu}^{2D-6}{\beta}^{2}}{ \left( D-3 \right) ^{\frac{D-5}{2}}
 \left(D-1 \right) ^{3}}} \Bigg{\}} \frac{{r}^{5-4\,D}
 {\nu}^{3-D}}{{\beta}^{2}}{dr}\nonumber \\
&& -\ln
 \left( \frac{D-3}{2} \right)  \left( 
\frac{16}{3}\,{\frac { \left( D-3 \right)
^{\frac{3D-7}{2}}}{ 
\left( D-1 \right) ^{\frac{3D-7}{2}} \left( D-2 \right) {
\nu}^{3\,D-9}}}+\frac{2}{3}\,{\frac { 
\left( D-3 \right) ^{\frac{3D-3}{2}}}{
 \left( D-1 \right) ^{
\frac{3D-5}{2}}{\beta}^{2}{\nu}^{3\,D-7}}} \right)
 \Bigg{\}}+O(q^{5}) \ ,
\end{eqnarray}
\begin{eqnarray}\label{aphiphi}
\hspace*{-2.5cm}
a_{\varphi}&=&\frac{-\nu q}{r^{D-3}}+
q^{3}r^{2}\Bigg{\{} \int \! \Bigg{\{} - \left( \frac{D-1}{2} \right)
 \left( \frac{D-3}{2} \right) ^{
\frac{D-5}{2}}S_{{5}}{r}^{3\,D-3}{\nu}^{D-3}
\int \!S_{{1}}S_{{5}}\int \!{
\frac {{r}^{D-2}}{{S_{{4}}}^{2}
\left( \left( \frac{D-3}{2}
\right) {r}^{2}- \left( \frac{D-1}{2} \right) {\nu}^{2}
 \right) ^{4}}}{dr}\nonumber \\
&& {r}^{2-3\,D}{dr}-\left(
\frac{D-1}{2} \right) S_{{2}}{\nu}^{D-3}S_{{4}}
\left( \left( \frac{D-3}{2}
 \right) {r}^{2}-\left( \frac{D-1}{2}
\right) {\nu}^{2} \right) ^{2}\int
\!{\frac {{r}^{D-2}}{{S_{{4}}}^{2}
\left( \left( \frac{D-3}{2} \right) {r}
^{2}- \left( \frac{D-1}{2} \right) {\nu}^{2}
\right) ^{4}}}{dr}\nonumber \\
&& \left( \frac{D-3}{2} \right) ^{\frac{1-D}{2}}-
{r}^{D-1}S_{{3}
} \Bigg{\}} \frac{1}{S_{{5}}{r}^{4\,D-3}{\beta}^{2}{\nu}^{2\,D-7}}
 {dr}+\frac{16}{3}\,{\frac {
\left( D-3
 \right) ^{\frac{3D-5}{2}}
\left( D-1 \right) ^{\frac{5-3D}{2}}\ln
\left( \frac{D-3}{2} \right) {\nu}^{8-3\,D}}{D-2}}\nonumber \\
&&+\frac{2}{3}\,{\frac { \left( D-3
 \right) ^{\frac{3D-1}{2}}
\left( D-1 \right) ^{\frac{3-3D}{2}}\ln
\left(\frac{D-3}{2}
\right) {\nu}^{-3\,D+6}}{{\beta}^{2}}}\Bigg{\}}+O(q^{5}) \ ,
\end{eqnarray}
where in the above
equations $S_{1}$, $S_{2}$, $S_{3}$, $S_{4}$,
and $S_{5}$ are
\begin{eqnarray}\label{S1}
S_{1} &=& \left( - \left( D-2 \right)
\left( D-3 \right) ^{\frac{D+5}{2}
}{2}^{\frac{3-D}{2}}{\nu}^{D-3}-
{\frac { \left( D-3 \right) ^{\frac{D+1}{2}
}{2}^{\frac{9-D}{2}}{\beta}^{2}
\left( D-1 \right) {\nu}^{D-1}}{D-2}}
 \right) {r}^{D+1}\nonumber \\
&&+ \left( D-1 \right) ^{3} \left( D-3
\right) ^{\frac{D+1}{2}}{2}^{
\frac{3-D}{2}}{r}^{D-1}{\nu}^{D-1}+3\, \left( D-2 \right)
 \left( D-3 \right) ^{4}
\left( D-1 \right) ^{\frac{D-1}{2}}{2}^{
\frac{1-D}{2}}{r}^{4}{\nu}^{2\,D-6}\nonumber \\
&&- \left(  \left( D-2 \right) ^{2}+
\frac{D-1}{2}\right)  \left( 3\,D-7
\right)  \left( D-3 \right) ^{2} \left( D-1
 \right) ^{\frac{D-1}{2}}{2}^{\frac{3-D}{2}}{r}^{2}{\nu}^{2
\,D-4}\nonumber \\
&&+ \left(3\,D-5 \right)  \left( D-3 \right) ^{2}
\left( D-1 \right) ^{\frac{D+3}{2}}{2}^{
\frac{1-D}{2}}{\nu}^{2\,D-2} \, ,
\end{eqnarray}
\begin{eqnarray}\label{S2}
S_{2}&=& \left( 64\,{\frac { \left( D-3
\right) ^{\frac{3D-7}{2}}{\nu}^{2
}{\beta}^{2}}{ \left( D-1 \right) ^{
\frac{D-7}{2}}{2}^{D+2} \left( D-2
 \right) ^{2}}}-\frac{16}{3}\,{\frac {
\left( D-3 \right) ^{\frac{1+3D}{2}}}{
 \left( 3\,D-7 \right)  \left( D-1 \right) ^{
\frac{D-3}{2}} \left(3\,
D-5 \right) }} \right) {r}^{3\,D-3}\nonumber \\
&&-16\, \left( D-3 \right) ^{D-2}{2}^{2
-D} \left( \frac{1}{8}\, \left( D-3 \right) ^{2}
\left( D-2 \right) {\nu}^{D-3
}+{\frac {{\beta}^{2} \left( D-1
\right) {\nu}^{D-1}}{D-2}} \right) {r
}^{2\,D}\nonumber \\
&&+8\, \left( D-1 \right) ^{\frac{D-1}{2}}{2}^{1-D}
 \left( D-3 \right) ^{\frac{D-1}{2}}
\left( 1/2\, \left( D-2 \right)
 \left( D-3 \right) ^{2}{\nu}^{2\,D-6}+{
\frac {{\beta}^{2} \left( D-1
 \right) {\nu}^{2\,D-4}}{D-2}} \right) {r}^{D+3}\nonumber \\
&&-32\, \left( D-1
 \right) ^{\frac{D-1}{2}}{2}^{-D-1} \left( D-3
\right) ^{\frac{D+1}{2}}
 \left(  \left( {\frac {17}{4}}-{\frac {15}{4}}
\,D+{D}^{2} \right) {
\nu}^{2\,D-4}+{\frac {{\beta}^{2} \left( D-1
\right) {\nu}^{2\,D-2}}{
 \left( D-2 \right) ^{2}}} \right) {r}^{D+1}
\nonumber \\
&&+\frac{8}{{2}^{D+1}}\, \left( 2\,D-3
 \right)  \left( D-3 \right) ^{\frac{D+1}{2}}\left( D-1
 \right) ^{\frac{D+1}{2}}{r}^{D-1}{\nu}^{2\,D-2}-
\left( D-2 \right)
 \left( D-3 \right) ^{3} \left( \frac{D-1}{2}
\right) ^{D-1}{r}^{6}{\nu
}^{3\,D-9}\nonumber \\
&&+4\, \left( D-3 \right) ^{2} \left( \frac{D-1}{2}
\right) ^{D-1} \left( -{\frac {103}{4}}+
{\frac {133}{4}}\,D-{\frac {59}{4}}\,{D}^
{2}+\frac{9}{4}\,{D}^{3}
\right) {r}^{4}{\nu}^{3\,D-7} \left( 3\,D-7 \right) ^
{-1}\nonumber \\
&&-8\, \left( {\frac {27}{8}}-\frac{7}{2}\,D+
{\frac {9}{8}}\,{D}^{2}
 \right)  \left( D-3 \right) ^{2} \left( D-2
\right)  \left( D-1
 \right) ^{D-1}{2}^{1-D}{r}^{2}{\nu}^{3\,D-5}
\left( 3\,D-5 \right)
^{-1}\nonumber \\
&&+\frac{2}{3}\, \left( D-3 \right) ^{2}
\left( 3\,D-5 \right)  \left( D-1
 \right) ^{D}{2}^{-D}{\nu}^{3\,D-3}+4\,
\left( D-1 \right) ^{2} \left( D-3 \right) ^{D-1}{2}^{1-D
}{r}^{2\,D-2}{\nu}^{D-1}\,,
\end{eqnarray}
\begin{eqnarray}\label{S3}
S_{3} &=&-{r}^{D+1}{\nu}^{2\,D-6}{D}^{2}-{
\frac { \left( -235+37\,D+2\,
 \left( D-11 \right) ^{2} \right)  \left( D-1
\right) ^{\frac{D-1}{2}}
 \left( D-3 \right) ^{
\frac{-D+5}{2}}{r}^{2}{\nu}^{3\,D-7}}{3\,D-5}}\nonumber \\
&&+{
\frac { \left( D-2 \right)
\left( D-1 \right) ^{\frac{D-1}{2}} \left( D-
3 \right) ^{\frac{-D+7}{2}}{r}^{4}{
\nu}^{3\,D-9}}{3\,D-7}}+{r}^{3\,D-3}
 \left( D-3 \right) ^{\frac{-D+3}{2}}
\left( D-1 \right) ^{\frac{D-1}{2}}{
\nu}^{3\,D-9}{\beta}^{2}\nonumber \\
&&+ \left( D-3 \right)  \left( D-1 \right) {r}^{
D-1}{\nu}^{2\,D-4}-9\,{r}^{D+1}{\nu}^{2
\,D-6}+6\,{r}^{D+1}{\nu}^{2\,D-
6} \left( D \right) -4\,{
\frac {{r}^{D+1}{\beta}^{2}{\nu}^{2\,D-4}{D}^
{2}}{ \left( D-2 \right) ^{2}
\left( D-3 \right) }}\nonumber \\
&&+16\,{\frac {{r}^{D
+1}{\beta}^{2}{\nu}^{2\,D-4}
\left( D \right) }{ \left( D-2 \right) ^{
2} \left( D-3 \right) }}-8\,{
\frac {{r}^{3\,D-3} \left( D-3 \right) ^{
D-3}{\beta}^{2} }{\left( D-1
\right) ^{D}(D-2)}}+4\,{\frac {{r}^{2\,D-2
} \left( D-3 \right) ^{\frac{D-3}{2}} {
\beta}^{2}{\nu}^{D-1}}{\left( D-1
\right) ^{\frac{D-5}{2}}(D-2)}}\nonumber \\
&&-4\,{\frac {{r}^{3\,D-3} \left( D-3
 \right) ^{D-3}{\beta}^{2} \left( D-1
\right) ^{-D}}{ \left( D-2
 \right) ^{2}}}-{\frac {27}{2}}
\,{r}^{2\,D} \left( D-1 \right) ^{D-1}{
\nu}^{4\,D-12}{\beta}^{2} \left( D-3
\right) ^{-D} \left( D \right)\nonumber \\
&& +{
\frac {28}{3}}\,{\frac {{r}^{3\,D-3}
\left( D-3 \right) ^{D} \left( D-
1 \right) ^{-D}{2}^{D} \left( D
\right) }{{\nu}^{2} \left( 3\,D-7
 \right)  \left( -5+3\,D \right) }}
+{\frac {27}{2}}\,{r}^{2\,D}
 \left( D-1 \right) ^{D-1}{\nu}^{4
\,D-12}{\beta}^{2} \left( D-3
 \right) ^{-D}\nonumber \\
&&+\frac{9}{2}\,{r}^{2\,D} \left(
D-1 \right) ^{D-1}{\nu}^{4\,D-12
}{\beta}^{2} \left( D-3
\right) ^{-D}{D}^{2}-24\,{\frac {{r}^{3\,D-3}
 \left( D-3 \right) ^{D-3}{\beta}^{2} {D}^{2}}
{\left( D-1 \right) ^{D}(D-2)}}
\nonumber \\
&&+{\frac {27}{2}}\,{r}^{2\,D-2}
\left( D-1 \right) ^{D-1}{\nu}^{4
\,D-10}{\beta}^{2} \left( D-3
\right) ^{-D} \left( D \right) +16\,{
\frac {{r}^{3\,D-3} \left( D-3
\right) ^{D-3}{\beta}^{2} {D}^{3}}{\left( D-1
 \right) ^{D} \left( D-2 \right) ^{2}}}\nonumber \\
&&+\frac{1}{3}\, \frac{\left( D-1
 \right) ^{\frac{D+1}{2}} {\nu}^{3\,D-5
}}{\left( D-3 \right) ^{\frac{D-5}{2}}}-
{\frac {12\,{r}^{D+1}{\beta}^{2}{\nu}^{2\,D-4}}{
\left( D-2 \right)
^{2} \left( D-3 \right) }}-\frac{9}{2}\,
\frac{{r}^{2\,D-2} \left( D-1
\right) ^{D-1}
{\nu}^{4\,D-10}{\beta}^{2} {D}^{2}}{
\left( D-3 \right) ^{D}}\nonumber \\
&&+\frac{1}{2}\,{r}^{2
\,D-2} \left( D-1 \right) ^{D-1}{\nu}^{4\,D-10}{\beta}^{2}
\left( D-3
 \right) ^{-D}{D}^{3}-4\,{\frac {{r}^{2\,D} \left( D-3
\right) ^{\frac{D-3}{2}} {\beta}^{2}{\nu}^{D-3}{D}^{2}}
{\left( D-1 \right) ^{\frac{D-1}{2}}(D-2)}}\nonumber \\
&&-\frac{1}{2}
\,{r}^{2\,D} \left( D-1
\right) ^{D-1}{\nu}^{4\,D-12}{\beta}^{2}
 \left( D-3 \right) ^{-D}{D}^{3}+16\,{
\frac {{r}^{3\,D-3} \left( D-3
 \right) ^{D-3}{\beta}^{2}  \left( D \right) }
{\left( D-1 \right) ^{D} \left( D-2 \right) ^{2}}}\nonumber \\
&&+24\,{\frac {{r}^{3\,D-3} \left( D-3
 \right) ^{D-3}{\beta}^{2}
\left( D-1 \right) ^{-D} \left( D \right) }
{D-2}}+\frac{4}{3}\,{\frac {{r}^{3
\,D-3} \left( D-3 \right) ^{D} \left( D-1
 \right) ^{-D}{2}^{D}{D}^{3}}{{\nu}^{2}
\left( 3\,D-7 \right)  \left(
-5+3\,D \right) }}\nonumber \\
&&-4\,{\frac {{r}^{3\,D-3} \left( D-3
\right) ^{D-3}{
\beta}^{2} \left( D-1 \right) ^{-D}{D}^{4}}{ \left( D-2 \right) ^{2}}}
-24\,{\frac {{r}^{3\,D-3} \left( D-3
\right) ^{D-3}{\beta}^{2} \left(
D-1 \right) ^{-D}{D}^{2}}{ \left( D-2
\right) ^{2}}}\nonumber \\
&&-12\,{\frac {{r}^{
2\,D} \left( D-3 \right) ^{\frac{D-3}{2}} \left( D-1
\right) ^{\frac{-D+1}{2}}{\beta}^{2}{\nu}^{D-3}}{D-2}}+16
\,{\frac {{r}^{2\,D} \left( D-3
 \right) ^{\frac{D-3}{2}} {\beta}^{2}{
\nu}^{D-3} \left( D \right) }{\left( D-1
\right) ^{\frac{D-1}{2}}(D-2)}}\nonumber \\
&&-4\,{\frac {{r}^{3\,D-3} \left( D-3
\right) ^{D} {2}^{D}}{\left( D-1
 \right) ^{D}{\nu}^{2} \left( 3\,D-7
\right)  \left(3\,D-5
 \right) }}-{\frac {20}{3}}\,{\frac {{r}^{3\,D-3} \left( D-3 \right) ^
{D} {2}^{D}{D}^{2}}{\left( D-1 \right) ^{D}{\nu}^{2} \left( 3\,D-7
 \right)  \left( 3\,D-5 \right) }}
\nonumber \\
&&+{\frac {8{r}^{3\,D-3} \left( D-3
\right) ^{D-3}{\beta}^{2} {D}^{3}}{\left( D-1
\right) ^{D}(D-2)}}-{\frac {27}{2}}\,{r}^{2\,D-2}
\left( D-1 \right) ^{D-1}{\nu}^{4\,D-10}{\beta}^{2}
\left( D-3
\right) ^{-D}\,,
\end{eqnarray}
\begin{eqnarray}\label{S4}
S_{4}&=&\sum^{\frac{D-5}{2}}_{i=0}(i+
1)\nu^{2i}(\frac{D-1}{2})^{i}(
\frac{D-3}{2})^{\frac{D-2i-7}{
2}}r^{D-2i-5}\,,
\end{eqnarray}
\begin{eqnarray}\label{S5}
S_{5}&=&{r}^{D-1}-\frac{1}{2}\,{\frac { \left( D-1
\right) ^{\frac{D-1}{2}}{\nu}^
{D-3}{r}^{2}}{ \left( D-3 \right) ^{\frac{D-3}{2}}}}+
\frac{1}{2}\,{\frac {
\left( D-1 \right) ^{\frac{D-1}{2}} {\nu}^{2D-4}}{
\left( D-3 \right) ^{\frac{D-3}{2}}{\nu}^{D-3}}}\,.
\end{eqnarray}

\acknowledgments{The support of the Funda\c{c}\~{a}o para a
Ci\^{e}ncia e a Tecnologia (FCT) of Portugal, Projects
PTDC/FIS/098962/2008 and PEst-OE/FIS/UI0099/2011 is gratefully
acknowledged.  M.A. is supported by an FCT grant.  The works of 
S.H.H. and A.S. have been financially supported by the Center for
Excellence in Astronomy and Astrophysics of IRAN (CEAAI-RIAAM).

\end{document}